\pdfoutput=1
\documentclass[journal]{IEEEtran}

\usepackage{cite}

\ifCLASSINFOpdf
  \usepackage[pdftex]{graphicx}

  \DeclareGraphicsExtensions{.pdf,.jpeg,.png}
\else

\fi
\ifCLASSOPTIONcompsoc
\usepackage[caption=false,font=normalsize,labelfon
t=sf,textfont=sf]{subfig}
\else
\usepackage[caption=false,font=footnotesize,subrefformat=parens,labelformat=parens]{subfig}
\fi
\usepackage{amsmath}
\usepackage{stix}
\usepackage{threeparttable}
\usepackage{array}
\usepackage{xcolor}
\usepackage{soul}
\soulregister\cite7

\usepackage{pdfpages}
\usepackage{tikz}
\newcommand*\circled[1]{\tikz[baseline=(char.base)]{
    \node[shape=circle,draw,inner sep=0.5pt] (char) {\footnotesize#1};}}
\usetikzlibrary{calc,shapes,arrows,chains,arrows.meta}
\usepackage[export]{adjustbox}
\usepackage{upgreek}
\usepackage{makecell}
\usepackage{multirow}
\usepackage{placeins}
\definecolor{newTextColor}{RGB}{0,0,0} 
\definecolor{modTextColor}{RGB}{0,0,0} 
\definecolor{typoColor}{RGB}{0,0,0}

\newcounter{node}
\usepackage[hidelinks]{hyperref}
\makeatletter
\newcommand{\customlabel}[2]{%
  \protected@write \@auxout {}{\string \newlabel {#1}{{#2}{\thepage}{#2}{#1}{}}}%
  \hypertarget{#1}{#2}
}
\tikzset{autonumbered node/.style={/utils/exec={\stepcounter{node}}},label=above:{\alph{node}}}

\makeatother

\begin{document}

\onecolumn
\pagenumbering{Roman}
\textbf{Author's pre-print}
\newline\newline
$\copyright$ 2024 IEEE. Personal use of this material is permitted. Permission
from IEEE must be obtained for all other uses, in any current or future
media, including reprinting/republishing this material for advertising or
promotional purposes, creating new collective works, for resale or
redistribution to servers or lists, or reuse of any copyrighted
component of this work in other works.

This article has been accepted for publication in \textit{IEEE Antennas and Wireless Propagation Letters}. This is the author's version which has not been fully edited and
content may change prior to final publication. 
Citation information: DOI~10.1109/LAWP.2024.3519709
\newpage
\twocolumn
\pagenumbering{arabic} 
%
\title{A Cupola-Shaped Multimode Multiport Antenna \\ for Aerial Direction Finding}
%
%
%
\author{Lukas~Grundmann, \IEEEmembership{Graduate Student Member, IEEE}, Marc Gerlach, Wolfgang Schäfer and  Dirk~Manteuffel, \IEEEmembership{Member, IEEE}

\thanks{This work is performed in the project Master360 under grant 20D1905C, funded by the German Federal Ministry for Economic Affairs and Climate Action within the Luftfahrtforschungsprogramm (LuFo). The authors would also like to thank Sami A. Almasri, Nils L. Johannsen and Peter A. Hoeher from Kiel University for preparing the DF algorithm used in the flight test.
}
\thanks{Lukas Grundmann and Dirk Manteuffel are with the Institute of Microwave and Wireless Systems, Leibniz University Hannover, Appelstr. 9A, 30167 Hannover, Germany \mbox{(e-mail:} \mbox{grundmann@imw.uni-hannover.de;} \mbox{manteuffel@imw.uni-hannover.de)}
}
\thanks{Marc Gerlach and Wolfgang Schäfer are with f.u.n.k.e. AVIONICS GmbH, Magirusstr. 39/1, 89077 Ulm, Germany \mbox{(e-mail:} \mbox{marc.gerlach@funkeavionics.de;} \mbox{wolfgang.schaefer@funkeavionics.de)}
}}

\maketitle

\begin{abstract}
A multimode multiport antenna (M³PA), constisting of a single conducting structure, is proposed for use with the airborne collision avoidance system (ACAS). Its cupola-shaped surface is selected from a set of aerodynamically suitable and symmetric structures. The selection is based on a presented statistical evaluation procedure for direction-finding (DF) antennas and algorithms, which predicts the root mean square error (RMSE) of the estimation using characteristic modes (CMs). A demonstrator antenna and feed network is manufactured and its three ports are measured in an anechoic chamber. The results match the intended far-fields of an electric monopole and two magnetic dipoles. The presented flight test results show that the antenna is capable of unambiguous DF across the desired angular range in a real world environment. 
\end{abstract}


%
\IEEEpeerreviewmaketitle

\section{Introduction} \label{sec:Introduction}

\IEEEPARstart{T}{he} automated localization of air traffic participants by an unmanned aerial vehicle (UAV) is becoming increasingly important due to the rising popularity of UAVs \cite{Bauranov2021}. For cooperative aircraft, the airborne collision avoidance system (ACAS) constitutes the standard for this application. With the recent addition of ACAS-Xu \cite{ACASXu}, specific adaptations were made with respect to UAVs. 

The resulting integration conditions, in particular on small UAVs, also allow a new perspective on antenna development for aerial direction-finding (DF) applications \cite{Ghaemi2018}. 
Traditionally, an antenna array is installed on the ownship to determine the direction of arrival (DoA) of an electromagnetic signal, emitted by an aircraft (intruder) in proximity to the ownship~\cite{Pallavi2021}. 
However, multimode multiport antennas (M$^3$PAs), implemented as a single conducting structure, recently received increased attention for DF \cite{Ma2019, Poehlmann2019, Duplouy2019, Ren2021, Grundmann2023}. 
This concept provides a greater variety of antenna shapes, which can be utilized to simplify the integration of the antenna beneath a radome on an aircraft. 
A key concept in the design of M$^3$PAs are characteristic modes (CMs) \cite{Harrington1971a}. As eigenfunctions of the antenna structure, these are independent of the excitation. Critically, CMs can be utilized to predict the far-field properties of mutually orthogonal antenna ports by using the procedure described in \cite{Peitzmeier2022}. 

In our previous work \cite{Grundmann2023}, we introduced a deterministic procedure to evaluate M$^3$PA concepts in various stages of the design process by using CMs. 
While the antenna proposed in \cite{Grundmann2023} demonstrated the general feasibility of the evaluation and design process, {\color{modTextColor}its shape is aerodynamically disadvantageous.}
Therefore, this letter proposes an M$^3$PA suited for real world direction finding of ACAS signals. {\color{modTextColor}To this end, an evaluation metric is proposed that includes the statistical behavior of the DF algorithm. This metric is compatible with that proposed in \cite{Grundmann2023}, although it does not rely on it.}

\section{Statistical Evaluation of DF Antennas and Algorithms} \label{sec:theory}
In order to evaluate the DF performance of an antenna, a parameter that describes the relation between the possible DoAs is required.
In this work, a set of $K=341$ DoAs with vertically ($\Theta$, see Fig.~\ref{fig:SelectedModes}) polarized signals is considered. These are evenly distributed across the upper hemisphere to ensure {\color{modTextColor}a defined behavior for each DoA}. The lower hemisphere is assumed to be shaded by the aircraft fuselage and covered by a second antenna system if needed. 

{\color{modTextColor}For a plane wave incident on a $P$-port DF antenna from the $k$th DoA, the voltages measured at each antenna port are combined in the steering vector $\mathbf{x}_k$. 
In a noise-free environment, $\mathbf{x}_k$ is proportionate to the realized gain of the ports in the $k$th DoA. 
In the presence of noise, however, a DF algorithm is required to estimate the DoA of the plane wave from the noisy steering vector $\tilde{\mathbf{x}}$. 
In this letter, the most popular and versatile modern DF algorithm \cite{Kellogg2007}, the multiple signal classification (MUSIC) \cite{Schmidt1986} is employed, and it is briefly introduced in the following.} 
It utilizes an estimate of the covariance matrix 
\begin{equation}
\tilde{\mathbf{R}}_\mathrm{x} = \mathbf{\tilde{x}\tilde{x}}^\mathrm{H} \,,
\end{equation}
{\color{modTextColor}where Gaussian noise with a known signal to noise ratio SNR is assumed.} 
Next, an eigenvalue decomposition is performed. The eigenvector belonging to the highest eigenvalue forms the one-dimensional signal subspace, while the other eigenvectors form the $(P-1)$-dimensional noise subspace $\mathbf{N}$. 
{\color{modTextColor}From this, the $k$th entry of the MUSIC spectrum $\mathbf{P}_\mathrm{MUSIC}$ is calculated as
\begin{equation} \label{eq:PMUSIC}
    P_{\mathrm{MUSIC},k} = \left( \mathbf{x}_k^\mathrm{H} \mathbf{N} \mathbf{N}^\mathrm{H}  \mathbf{x}_k \right)^{-1} \,.
\end{equation}
Equation~(\ref{eq:PMUSIC}) is evaluated for all $K$ candidate DoAs to obtain the complete spectrum.} 
The maximum in $\mathbf{P}_\mathrm{MUSIC}$ is then selected as the estimated DoA.

{\color{modTextColor}This process is repeated many times in a Monte Carlo simulation. For each repetition, a true DoA is randomly chosen from the set of $K$ uniformly distributed DoAs and a random noise signal is added to the respective steering vector to obtain $\tilde{\mathbf{x}}$. From the resulting estimated DoAs, a root mean square error (RMSE) is determined as the overall metric for the DF performance of the antenna.} 

To illustrate the dependency of the RMSE on the SNR, a set of $P$ artificial, idealized far-fields is introduced. Their realized gain is determined through the terms of a Fourier series as
\begin{equation} \label{eq:refPattern}
\mathbf{G}_p(\Theta,\Phi) = \mathrm{e}^{\mathrm{j} p \Phi} \mathbf{e}_\Theta \,,
\end{equation}
with $p$ an integer and $0\leq p < P$. By conducting the above procedure with $P=3$ of these far-fields, it is found that $\Phi$ is estimated with an {\color{modTextColor}$\mathrm{RMSE}_\Phi$} of less than 15° for an $\mathrm{SNR}=-10\,\mathrm{dB}$. Since this constitutes a reasonable accuracy in the context of ACAS, this setup provides a reference, albeit idealized, for the following antenna development. Consequently, $\mathrm{SNR}=-10\,\mathrm{dB}$ is utilized in section~\ref{sec:SelectionOfBaseStructure}.

\section{Antenna Design}
\subsection{Selection of Base Structure} \label{sec:SelectionOfBaseStructure}
\begin{figure}
    \color{modTextColor}
    \centering
    \includegraphics[]{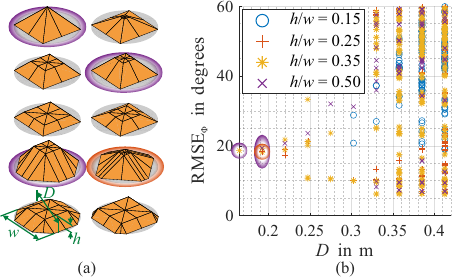}
    \caption{(a): Simulation models of the investigated conductive structures. Depictions are to scale. (b): $\mathrm{RMSE}_\Phi$ of sets of CMs of these structures, obtained through Monte Carlo simulations with 1000 runs per DoA.}
    \label{fig:StructureSelectionCombined}
\end{figure}
First, the basic geometric structure of the M$^3$PA is determined by applying the procedure from section~\ref{sec:theory} to the CM far-fields of various candidate structures, which are depicted in Fig.~\ref{fig:StructureSelectionCombined}(a). 
The investigated structures are positioned on a perfectly electrically conducting plane, which represents a simplified aircraft fuselage. Most intruder aircraft are expected to be near this plane. 
The structures possess a square symmetry, which simplifies the creation of ports in section~\ref{sec:MatchingOfPorts}. The first three rows of Fig.~\ref{fig:StructureSelectionCombined}(a) display pyramids or pyramid trunks, while the last two contain cupolas. 
Each structure is simulated at a frequency of 1090\,MHz with varying diameter of its smallest circumscribing sphere $D$. For every possible set of CMs {\color{modTextColor}of any of the structures that fulfills the following criteria, the $\mathrm{RMSE}_\Phi$} is determined and depicted in Fig.~\ref{fig:StructureSelectionCombined}(b): 
\begin{itemize}
    \item Set contains at least three modes
    \item Each mode can be assigned to a unique, independent port
    \item Degenerate modes only appear as complete sets
    \item Eigenvalue magnitude of all modes is below three
\end{itemize}
Here, a small CM eigenvalue magnitude $|\lambda_p|$ indicates a mode closer to resonance, which later simplifies port matching \cite{CabedoFabres2007}. Degenerate modes have the same eigenvalue. The common criterion of $|\lambda_p|\leq 1$ is extended here to allow a greater variety of modal sets, since the required bandwidth {\color{newTextColor}of $1090\,\mathrm{MHz}\pm1\,\mathrm{MHz}$} is relatively narrow. 

{\color{modTextColor}Five sets of modes in Fig.~\ref{fig:StructureSelectionCombined}(b) are of particular interest due to the small value of $D$. These are encircled in violet and red in Fig.~\ref{fig:StructureSelectionCombined}. 
Since all of them possess a similar $\mathrm{RMSE}_\Phi = 18.7^\circ \pm 0.6^\circ$, aerodynamic considerations are employed to select one of the corresponding structures.} %
{\color{newTextColor}For an aerodynamically streamlined radome, which is to cover the antenna in a practical application, a smaller height reduces the drag force \cite{Giragosian1990}. This favors the cupola encircled in red, as it has the smallest height to width ratio $h/w = 0.25$ and the lowest absolute height. 
Additionally, the cupola provides the lowest average cross sectional area, which is proportionate to the drag force. }%
{\color{modTextColor}Due to these aerodynamic advantages over the other structures from Fig.~\ref{fig:StructureSelectionCombined}(a), as well as the antenna from \cite{Grundmann2023}, the cupola is chosen as the base structure of the antenna.} The angle between its rectangular surface elements and the ground plane is 30°. 

\subsection{Matching of Ports} \label{sec:MatchingOfPorts}
\begin{figure}
    \centering
    \includegraphics[]{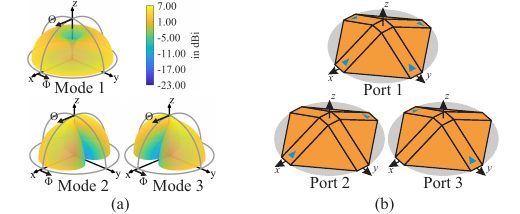}
    \caption{(a): $\Theta$-components of far-fields of the selected CMs 1, 2 and 3 at 1090\,MHz. (b): Basic feed setups of the corresponding ports in blue. The eigenvalues of the modes are $\lambda_1=-2.2$, $\lambda_2=2.5$ and $\lambda_3=2.5$.}
    \label{fig:SelectedModes}
    \label{fig:DeltaGaps}
\end{figure}
The $\Theta$-components of the far-field directivities of the selected CMs are depicted in Fig.~\ref{fig:SelectedModes}(a). Mode 1 possesses an electric monopole pattern, while modes 2 and 3 are degenerate and have magnetic dipole patterns. {\color{newTextColor}The objective is now to implement matched ports whose far-fields closely resemble those of the CMs.}

{\color{modTextColor}For each of the selected CMs, the respective port consists of a set of delta gap feeds, which are depicted in Fig.~\ref{fig:DeltaGaps}(b)}. Crucially, all three port arrangements utilize the same four delta gap feeds, but with different weights. They are derived using the symmetry properties of the antenna structure, as detailed in~\cite{Grundmann2023} {\color{newTextColor}and \cite{Peitzmeier2022}}. {\color{modTextColor}This way, the ports are uncorrelated from each other, and each port is correlated to one of the three selected modes.} 

\begin{figure}
    \centering
    \includegraphics[scale=0.8]{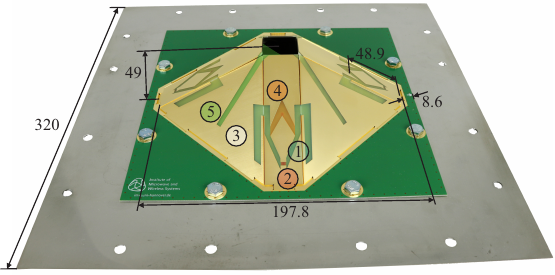}
    \caption{Photograph of the manufactured demonstrator antenna with important dimensions in mm. Regions of interest are highlighted in different colors and marked by corresponding encircled numbers. The antenna is bolted to a square steel plate.}
    \label{fig:AriesDimensionsJoined}
\end{figure}

{\color{modTextColor}Now, the delta gap feeds need to be replaced by practical coupling structures, which provide input matching for the ports over the relevant bandwidth. 
In the ACAS downlink band $1090\,\mathrm{MHz}\pm1\,\mathrm{MHz}$, all ports need to be matched. Since in the uplink band $1030\,\mathrm{MHz}\pm0.2\,\mathrm{MHz}$, no DF capability is required, matching port~1 is sufficient for transmitting omnidirectional interrogation signals. 
These requirements result in the manufactured antenna} depicted in Fig.~\ref{fig:AriesDimensionsJoined}, which is derived and explained in the following. 

\begin{figure}
    \centering
    \color{modTextColor}
    \includegraphics[scale=0.78]{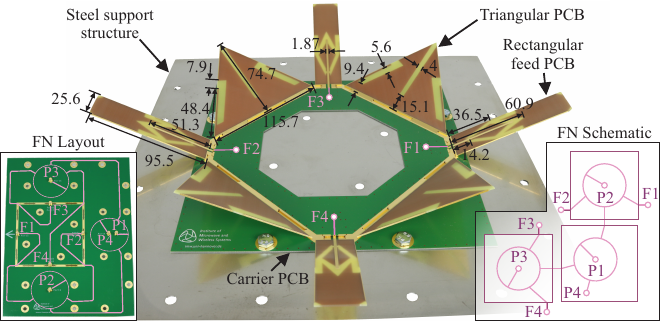}
    \caption{Photograph of the individual components of the demonstrator antenna with dimensions given in mm. Layout and schematic of the feed network FN are displayed in the bottom corners, where P stands for port and F for feed. Transmission lines are highlighted in pink. For the antenna to be used, the PCBs are folded inwards and soldered together, which yields the setup shown in Fig.~\ref{fig:AriesDimensionsJoined}.}
    \label{fig:AriesDimensionsSplit}
\end{figure}
The antenna is composed of nine printed circuit boards (PCBs), which are depicted in Fig.~\ref{fig:AriesDimensionsSplit}. All PCBs are made of 1\,mm thick FR-4 substrate. The carrier PCB is fully metalized on both sides and coated with green solder mask. {\color{modTextColor}The cutout in the center is covered by the steel support structure, to which the antenna is bolted. Furthermore, the carrier PCB possesses slots in which the other PCBs are inserted.} 
Four triangular PCBs, metalized only on the outside, are inserted here.
The four remaining slots are occupied by the rectangular feed PCBs. They are metalized on the outside of the antenna, while a microstrip line stub is positioned on the inside. 
On the carrier PCB, microstrip lines lead from the feed PCBs to four coaxial connectors below the antenna (not visible in Fig.~\ref{fig:AriesDimensionsSplit}). 

From the coaxial connectors, the received signals are distributed to the ports of the antenna with the help of a feed network. {\color{modTextColor}It is sketched in Fig.~\ref{fig:AriesDimensionsSplit}, a detailed layout is found in \cite{Grundmann2023}.} The feed network consists of three ring hybrid couplers. Due to this layout, a fourth port is created, which is terminated by a $50\,\Omega$ load during operation. The other three ports resemble the intended configurations from Fig.~\ref{fig:DeltaGaps}(b). 

To allow signals to couple to the microstrip lines on the feed PCBs, slots are introduced on the outer surface of the antenna. {\color{modTextColor}The shape and dimensions of these slots are designed to achieve input matching of the antenna ports over the desired bandwidth.} The length of the main slot (\circled{1} in Fig.~\ref{fig:AriesDimensionsJoined}) of 149\,mm is designed to roughly match half the free space wavelength at the operating frequency. 
In this slot, the main coupling from the microstrip line happens in the lower section of the rectangular feed PCB (\circled{2} in Fig.~\ref{fig:AriesDimensionsJoined}). 
{\color{modTextColor}Simulations showed that avoiding the center region of the triangular PCB with the main slot (\circled{3} in Fig.~\ref{fig:AriesDimensionsJoined}) assists in moving the resonance frequencies of port 1 and ports 2 and 3 closer together}. Therefore, the main slot only extends about 8\,mm onto the triangular PCBs, before it runs parallel to the edge of the rectangular PCB. 
An additional slot \circled{4} interrupts the direct conductive path from the feed point to the peak of the cupola, replacing it with a capacitive coupling. It is used to tune all ports of the antenna simultaneously. 

For the currents excited by ports 2 and 3, the distance covered when traveling across the peak of the original coupola is found to be too short. Therefore, the square segment at the peak of the antenna is removed and a slot is introduced on the triangluar PCBs (\circled{5} in Fig.~\ref{fig:AriesDimensionsJoined}), which blocks the direct path. Port 1 is not significantly affected by this, since the currents excited by it run tangential to the slot and possess a minimum at the peak of the cupola. 
Finally, the exact dimensions of all slots are determined in an iterative process of finite difference time domain (FDTD) simulations, conducted in EmpireXPU \cite{empirexpu}, resulting in the demonstrator depicted in Fig.~\ref{fig:AriesDimensionsJoined}. {\color{newTextColor}Its simulated far-fields closely match those of the CMs from Fig.~\ref{fig:SelectedModes}(a), with envelope correlation coefficients of $0.96$ for port~1 and $0.92$ for ports~2 and 3. }

\section{Measurement and Test}
\subsection{Lab Measurements} \label{sec:LabMeasurements}
\begin{figure}
    \centering
    \includegraphics[width=0.9\columnwidth]{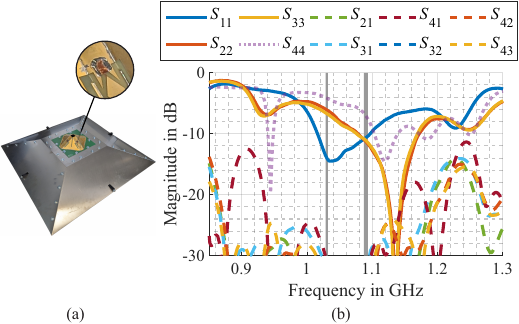}
    \caption{(a): Photograph of the measured and tested demonstrator antenna, positioned on a mock-up of the aircraft hull. (b): Scattering parameters of the proposed antenna, measured at the ports of the feed network. The two relevant frequency bands are indicated in grey. }
    \label{fig:LabMeasurementSParams}
\end{figure}
In order to perform measurements with the manufactured demonstrator, it is positioned on a mock-up of a generic aircraft hull, as can be seen in Fig.~\ref{fig:LabMeasurementSParams}(a). The mock-up is a steel truncated square pyramid with 100\,mm height and 800\,mm bottom side length. The feed network used in the measurements is positioned inside this mock-up and therefore not visible in Fig.~\ref{fig:LabMeasurementSParams}(a). 
A section of the tuning slot \circled{4} is covered with copper tape to optimize input matching, as shown in the magnification in Fig.~\ref{fig:LabMeasurementSParams}(a). This is required due to manufacturing inaccuracies. These are likely caused by imperfect soldering and deviations in the permittivity of the FR-4 substrate material. 
{\color{modTextColor}The measured scattering parameters are depicted in Fig.~\ref{fig:LabMeasurementSParams}(b). It is found that an input reflection of less than $-10$\,dB is achieved for the ports over the relevant bandwidths specified in section~\ref{sec:MatchingOfPorts}.}

\begin{figure}
    \centering
    \includegraphics[width=\columnwidth]{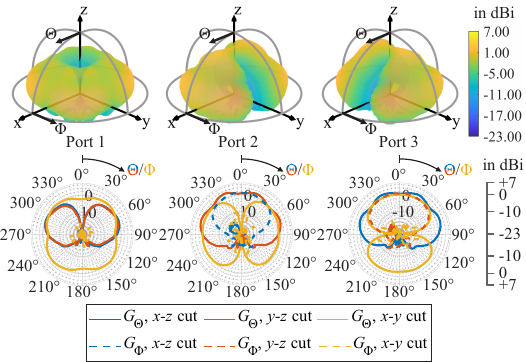}
    \caption{Measured realized gain of the ports 1, 2 and 3 of the proposed antenna depicted in Fig.~\ref{fig:LabMeasurementSParams}(b). Unless stated otherwise, the $\Theta$-component is depicted.}
    \label{fig:LabMeasurementFarFields}
\end{figure}
The far-field characteristics of the antenna are measured in an anechoic chamber. The measured realized gain of the three ports is given in Fig.~\ref{fig:LabMeasurementFarFields}. Overall, a reasonable agreement with the CM far-fields from Fig.~\ref{fig:SelectedModes} {\color{newTextColor}and a dominant $\Theta$-polarization are} observed, considering the difference in the utilized ground structures. Still, some small contributions by other modes are observable in the slightly indented shape of the $x$-$y$ cut of port 1 and the dips in the $y$-$z$ and $x$-$z$ cuts of ports 2 and 3, respectively. {\color{newTextColor}The total radiation efficiencies are affected by losses in the feed network, resulting in 56\%, 67\% and 64\% at 1090\,MHz for ports~1 to 3, respectively, and 63\% for port~1 at 1030\,MHz.}

\subsection{Flight Test}
\begin{figure}
    \centering
    \includegraphics[width=\columnwidth]{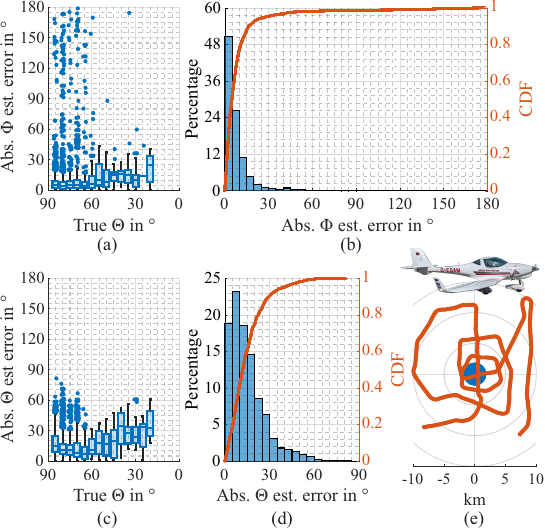}
    \caption{Results of flight tests. (a) and (c): Absolute  estimation errors of $\Phi$ and $\Theta$, respectively, depicted as box plots in dependency of the true $\Theta$ of the intruder. (b) and (d): respective histograms. (e): Photograph of the intruder aircraft and its actual flight path (red), relative to the position of the AUT (blue).}
    \label{fig:FlightTest}
\end{figure}
To demonstrate the DF capabilities of the developed antenna system, a flight test is conducted. {\color{modTextColor}To simplify the setup, the antenna under test (AUT) is stationary, positioned on the roof of a four-story building.} The antenna configuration from Fig.~\ref{fig:LabMeasurementSParams}(a) is used together with a custom build receiver and a Mode-S interrogator, which uses a separate monopole antenna positioned in close proximity to the AUT. For the flight test, the intruder aircraft, an Aquila~211, performed a pre-defined flight around the AUT in Ulm, Germany, on May 14th 2024. The true flight track is depicted in Fig.~\ref{fig:FlightTest}(e). The altitude of the intruder above ground varies between 850\,m and 950\,m. To test the performance for various distances and directions, the intruder flies pass and around the AUT at various distances, as well as over the AUT from different azimuth angles. 
An offset of $-8.6^\circ$ is corrected in post processing to set the average of the azimuth error to zero. This compensates the misalignment of the AUT with true north and other inaccuracies in the experimental setup. 

The otherwise unfiltered results of the flight test are depicted in Figs.~\ref{fig:FlightTest}(a)-(d). By comparing the histograms (b) and (d), it is evident that the estimation of $\Phi$ is more reliable than that of $\Theta$, which is to be expected based on the far-fields. The RMSE of the $\Theta$ estimation is $18.9^\circ$, while that of the $\Phi$ estimation is $20.4^\circ$. However, the latter is mainly due to a few outliers, which are easily filtered in a real world application. When excluding the $1.5\%$ $\Phi$ errors above $90^\circ$, the $\mathrm{RMSE}_\Phi$ reduces to $11.6^\circ$. The median errors of the $\Theta$ and $\Phi$ estimation are $12.0^\circ$ and $4.9^\circ$, respectively. 
Additionally, the dependencies of the $\Phi$ and $\Theta$ errors on the true elevation of the intruder are depicted as box plots in Figs.~\ref{fig:FlightTest}(a) and (c). Here, outliers, which are typically fewer than $1\%$ of the values, are marked by dots. 
It is found that the estimation provides reasonable results for all investigated elevation angles, but performance is best in the region where most real world intruders are expected, which is $90^\circ\leq\Theta\leq60^\circ$.

\section{Conclusion}
The simple setup of only three ports, used in the proposed aerial DF antenna, provides good DF performance in the flight test. In particular, the azimuth of intruders anywhere in the upper hemisphere is reliably estimated, in contrast to the antenna from \cite{Grundmann2023}. Furthermore, the antenna possesses a low profile {\color{modTextColor}and its shape enables an efficient aerodynamic integration.} 
In future applications, it might be considered to move the feed network to the digital domain, which is made possible due to the relatively well matched fourth port of the feed network. 

\ifCLASSOPTIONcaptionsoff
  \newpage
\fi



\bibliographystyle{IEEEtran}
\bibliography{IEEEabrv,literatur_master360}

\end{document}